\documentclass[journal]{IEEEtran}

\ifCLASSINFOpdf
\else
   \usepackage[dvips]{graphicx}
\fi
\usepackage{url}
\usepackage{amsmath}
\usepackage{caption}
\usepackage{graphicx}
\usepackage{multirow}
\usepackage{amssymb}
\usepackage{svg}

\begin{document}

\title{Mel-FullSubNet: Mel-Spectrogram Enhancement for Improving Both Speech Quality and ASR}

\author{Rui Zhou$^{\#}$, Xian Li$^{\#}$, Ying Fang and Xiaofei Li$^*$
\thanks{}
\thanks{ The authors are with Westlake University and also with Westlake Institute for Advanced Study, Hangzhou, China. $^{\#}$ equal contribution. * Correspondence: lixiaofei@westlake.edu.cn}
}

\markboth{Journal of \LaTeX\ Class Files, Vol. 14, No. 8, August 2015}
{Shell \MakeLowercase{\textit{et al.}}: Bare Demo of IEEEtran.cls for IEEE Journals}
\maketitle

\begin{abstract}
In this work, we propose Mel-FullSubNet, a single-channel Mel-spectrogram denoising and dereverberation network for improving both speech quality and automatic speech recognition (ASR) performance. Mel-FullSubNet takes as input the noisy and reverberant Mel-spectrogram and predicts the corresponding clean Mel-spectrogram. The enhanced Mel-spectrogram can be either transformed to  speech waveform with a neural vocoder or directly used for ASR. Mel-FullSubNet encapsulates interleaved full-band and sub-band networks, for learning the full-band spectral pattern of signals and the sub-band/narrow-band properties of signals, respectively. Compared to linear-frequency domain or time-domain speech enhancement, the major advantage of Mel-spectrogram enhancement is that Mel-frequency presents speech in a more compact way and thus is easier to learn, which will benefit both speech quality and ASR. Experimental results demonstrate a significant improvement in both speech quality and ASR performance achieved by the proposed model.
Code and audio examples of our model are available online \footnote{https://audio.westlake.edu.cn/Research/Mel-FullSubNet.html}.

\end{abstract}


\IEEEpeerreviewmaketitle

\section{Introduction}

\IEEEPARstart{T}his work studies single-channel speech enhancement using deep neural networks (DNNs), to improve both speech quality and automatic speech recognition (ASR) performance. 
A large class of speech enhancement methods employ DNNs to map from noisy and reverberant speech to corresponding clean speech, conducted either in time domain \cite{luo2019}, \cite{Defossez2020Real} or in time-frequency domain \cite{Xiong2022Spectro}, \cite{Hu2020DCCRN}, \cite{Andong2021Glance}. These methods can efficiently suppress noise, but not necessarily improve ASR performance due to the speech artifacts caused by speech enhancement networks \cite{iwamoto2022}. In \cite{kinoshita2020}, it was found that time-domain enhancement is more ASR-suitable than frequency-domain enhancement. In \cite{yang2022}\cite{nian2022}, different time-domain enhancement networks were developed for robust ASR. 


Mel-frequency models human frequency perception, within which speech enhancement is more perceptually and computationally efficient than within linear-frequency domain or time domain.
Sub-band networks \cite{Xiong2022Spectro}\cite{Li2020sub-band} and full-band/sub-band fusion networks, i.e. FullSubNet \cite{Fullsubnet}\cite{Zhou2023RTS}, have been recently proposed and achieved outstanding performance. However, separately processing sub-bands in the linear-frequency domain leads to a large computational complexity. To reduce the number of sub-bands and thus the complexity, Fast FullSubNet \cite{Fast fullsubnet} and the work of \cite{kothapally2023} proposed to perform sub-band processing in the Mel-frequency domain, and then transform to linear frequency with a joint post-processing network. In \cite{VoiceFixer}\cite{Diffusion}, speech enhancement is directly conducted in the Mel-frequency domain and then a separate neural vocoder is used to recover speech waveform, which achieve higher mean opinion scores compared to their linear-frequency counterparts. 
In this work, we propose Mel-FullSubNet, a single-channel Mel-spectrogram enhancement network for improving both speech quality and ASR performance, which is an extension of our previous full-band/sub-band fusion networks, i.e. FullSubNet \cite{Fullsubnet} and Fast-FullSubNet \cite{Fast fullsubnet}. The full-band network learns the spectral pattern of signals, and the sub-band network learns the sub-band properties of signals, such as the stationarity of signals and the convolutive relation of speech source and sub-band room filters. These information are complementary and thus their fusion improves the speech enhancement performance. Fast-FullSubNet \cite{Fast fullsubnet} minimizes the prediction error of enhanced linear-frequency spectrogram. By contrast, in this work, Mel-FullSubNet enhances Mel-spectrogram and minimizes the prediction error of enhanced Mel-spectrogram. The enhanced Mel-spectrogram can be directly used for ASR, or transformed to speech waveform using a separate neural vocoder as is done in \cite{VoiceFixer}. The network architecture of Mel-FullSubNet is revised over the one of FullSubNet and Fast-FullSubNet. The latter cascades  full-band and sub-band processing, while Mel-FullSubNet interleaves full-band and sub-band processing to reinforce the interaction between them. In addition, an online neural vocoder is developed in this work to enable online speech enhancement.  

Compared to linear-frequency spectrogram or time-domain waveform, Mel-frequency presents speech in a more compact way (but still perceptually efficient) and has a lower feature dimension (number of frequencies) from the perspective of machine learning, as a result the prediction error of enhancing Mel-spectrogram would be lower. This is beneficial for both speech quality and ASR: (i) Neural vocoders have been extensively studied in the field of Text-to-Speech, and are capable of efficiently transforming Mel-spectrogram back to time-domain waveform. Therefore, the low-error property of Mel-spectrogram enhancement can be maintained by the neural vocoders and higher speech quality can be achieved. (ii) Enhancing full-band speech needs to recover the full-band speech details and then compress to Mel frequency for ASR, which will obviously obtain less accurate Mel-spectrogram than directly enhancing Mel-spectrogram. More accurate Mel-spectrogram would be certainly beneficial for ASR.

\section{The proposed Method}
\begin{figure}[t]
  \centering
  \includegraphics[width=0.80\linewidth]{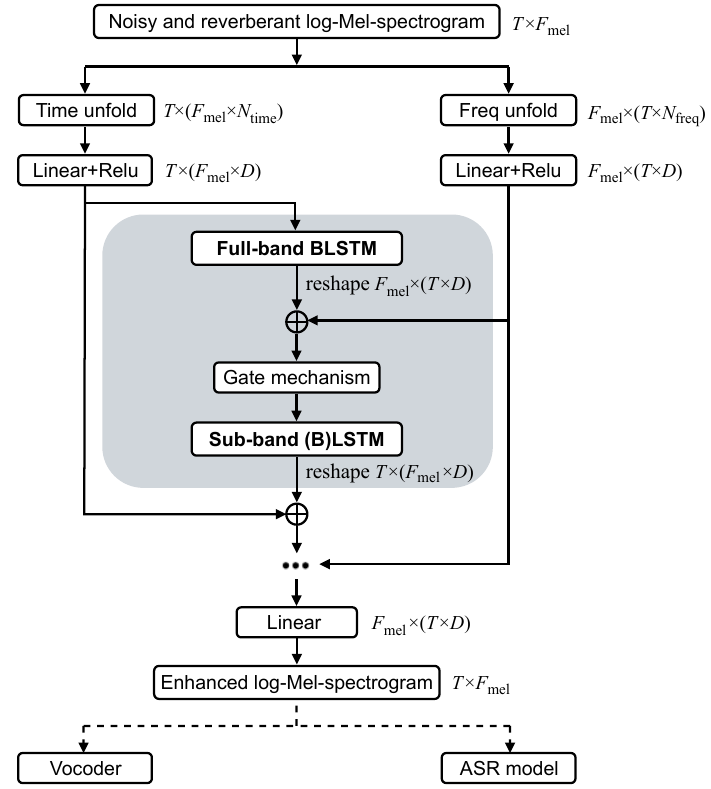}  
  \caption{\footnotesize{The proposed Mel-FullSubNet. Data dimension is presented in the format of:
number of sequences × (sequence length × feature dimension).}}
  \label{fig:net}
  \vspace{-0.5cm}
\end{figure}

The proposed single-channel Mel-spectrogram enhancement network is depicted in Fig.~\ref{fig:net}. 

\vspace{-0.3cm}

\subsection{Mel-FullSubNet} 

We denote the network input, i.e. log-Mel-spectrogram of single-channel noisy and reverberant speech recording, as ${Y}(t,f_{\text{mel}})$, and the corresponding clean log-Mel-spectrogram as ${X}(t,f_{\text{mel}})$, where $t\in\{1, \cdots, T\}$ and $f_{\text{mel}} \in\{1, \cdots, F_{\text{mel}}\}$ are the indices of time frame and Mel frequency, respectively. Data normalization methods will be introduced later respectively for speech enhancement and ASR. 
The spectrogram is reorganized in two ways to form the along-frequency and along-time sequences, which will be processed by the following full-band and sub-band LSTM networks, respectively. The along-frequency sequence is formed for each time frame, and there are a total of $T$ $F_{\text{mel}}$-long sequences. The input vector of each frequency step concatenates the signal component of one time frame and its adjacent $N_{\text{time}}-1$ time frames, which is then transformed to a $D$-dim input vector with a Linear layer. The along-time sequence is formed for each mel frequency, and there are a total of $F_{\text{mel}}$ $T$-long sequences. The input vector of each time step concatenates the signal component of one mel frequency and its adjacent $N_{\text{freq}}-1$ mel frequencies, which is also transformed to a $D$-dim input vector with a Linear layer.

Mel-FullSubNet is composed of multiple interleaved full-band and sub-band (B)LSTM networks. In between one full-band network and its following sub-band network, a gate mechanism is used. The vector dimension for representing each T-F bin is always $D$ throughout the network. After the final sub-band layer, one Linear layer transforms D-dim to 1-dim, resulting in the enhanced log-Mel-spectrogram $\hat{X}(t,f_{\text{mel}})$. The network is trained with the mean squared error (MSE) loss of enhanced log-Mel-spectrogram. 

The full-band networks process the $T$ along-frequency sequences independently. The first full-band layer processes the input along-frequency sequences, while other full-band layers take as input the summation of input sequences and (reshaped) sequences output by preceding sub-band layers. The full-band networks run recurrence along frequency to learn the full-band spectral pattern of signals to discriminate between speech and noise/reverberation. The adjacent time frames in input vectors provide some temporal context of spectral patterns. 
The sub-band networks process the $F_{\text{mel}}$ along-time sub-band sequences independently. All sub-band layers take as input the summation of input sequences and (reshaped) sequences output by preceding full-band layers. The adjacent frequencies in input vectors provide some frequency context for predicting the clean speech of one frequency. 
The sub-band networks learn the sub-band spectral pattern of signals and the sub-band/narrow-band properties of signals, such as learn the stationarity of signals to discriminate between non-stationary speech and stationary noise, and learn the convolutive relation between speech source and sub-band room impulse response to suppress reverberation.  The sub-band networks are set to be uni-directional LSTM for online processing, while bidirectional LSTM for offline processing. 

\subsection{Backends}
Mel-FullSubNet is followed by either a neural vocoder or an ASR model, but it is configured differently and  trained respectively when used for speech enhancement and ASR.
\subsubsection{\textbf{Neural Vocoder}}
\label{sec:vocos}
 The vocoder we adopt in this work is Vocos \cite{Vocos}, a recently proposed  Generative Adversarial Networks (GANs)-based neural vocoder. The generator of Vocos predicts the STFT coefficients of speech at frame level and then generates waveform through inverse STFT. Vocos uses the multiple-discriminators and multiple-losses proposed in HiFi-GAN \cite{kong2020hifi}, but it significantly improves the computational efficiency compared to HiFi-GAN that directly generates waveform at sample level. The original Vocos is designed for offline processing, as it employs non-causal convolution layers. To enable online processing, we modified Vocos to be causal by substituting every non-causal convolution layers with their causal version. As far as we know, this is the first time to try online neural vocoder in the filed. Surprisingly, the online vocoder still performs quite well.  The original Vocos and the modified online Vocos are used to work with the offline and online Mel-FullSubNet, respectively.

Directly feeding the enhanced Mel-spectrogram to Vocos pre-trained with clean speech yields undesired sound artifacts. Therefore, we decided to first pre-train the Mel-FullSubNet, and then jointly train it with a randomly initialized Vocos, which efficiently suppress the artifacts. 

In Vocos, time domain speech signal is normalized with a random gain to ensure that the maximum level of the resulting signal lies within the range of -1 to -6 dBFS. For offline speech enhancement, we apply this normalization to noisy signals and apply the same gain to corresponding clean signals, which are then used for training Mel-FullSubNet individually or jointly with Vocos. As for online speech enhancement, we apply a two-step normalization: (i) the utterance-level normalization is still applied as for the offline case to ensure a proper signal level for Vocos training. Then, a dataset-level global mean of noisy log-Mel-spectrogram is computed, denoted as $M$; (ii) an online normalization on log-Mel-spectrogram is applied as $\tilde{Y}(t,f_{\text{mel}})=Y(t,f_{\text{mel}}) - \mu(t) + M$ and $\tilde{X}(t,f_{\text{mel}})=X(t,f_{\text{mel}}) - \mu(t) + M$, where $\mu(t)$ is a recursively-calculated mean value of $Y(t,f_{\text{mel}})$, i.e. $\mu(t)=\alpha \mu(t-1)+(1- \alpha) \frac{1}{F} \sum_{f=1}^{F} {Y}(t,f_{\text{mel}})$. This online normalization transforms the mean of noisy log-Mel-spectrogram to a constant level, i.e. $M$, which facilitates the training of Mel-FullSubNet. The output of Mel-FullSubNet (namely the input of Vocos) is at the level of $M$, which is still consistent with the signal level of Vocos output, thence Mel-FullSubNet and Vocos can be jointly trained. At inference, only the online normalization is applied, and the joint network automatically outputs enhanced speech at the level of clean speech for training. The smoothing weight is set to $\alpha=\frac{L-1} {L+1}$, then recursive smoothing with $\alpha$ is equivalent to using a $L$-long rectangle smoothing window.   

\subsubsection{\textbf{ASR}}
When used for ASR, the proposed Mel-FullSubNet is independently trained, and the enhanced log-Mel-spectrogram is directly feed to an already-trained ASR system, without performing any fine-tuning or joint-training.  
Different ASR systems may have different configurations in STFT settings, number of Mel frequencies and base of logarithm. 
To seamlessly integrate the enhanced log-Mel-spectrogram into one ASR model, our Mel-FullSubNet would adopt the same configurations as the ASR model.   
A large portion of ASR systems apply per-utterance per-frequency mean normalization to the input log-Mel-Spectrogram. To better suit the ASR systems, the proposed Mel-FullSubNet also adopts this normalization.
The training cost of Mel-FullSubNet is not very high, so it can be easily re-trained for a new ASR system, especially for those large-scale ASR systems and already-deployed ASR systems. 


\section{Experiments}

\subsection{Experimental Configurations}
\textbf{Datasets} The proposed method was trained using the DNS challenge (INTERSPEECH 2020) dataset \cite{dns2020}. 
75\% of the clean speech clips were convolved with randomly selected room impulse responses (RIRs) from the Multichannel Impulse Response Database \cite{hadad2014}
and the training set of REVERB Challenge dataset \cite{reverb}. 
Speech and noise are mixed with a random signal-to-noise ratio (SNR) between -5 dB and 20 dB. 
The training target of Mel-FullSubNet is set as the direct-path speech or original clean speech. {Using this dataset, we have trained an online and an offline Mel-FullSubNet+Vocos for speech enhancement, and an offline Mel-FullSubNet for ASR.} 

The trained models are evaluated on three datasets:  (1) the test set of DNS, 
(2) the one-channel test set of REVERB, 
(3) the CHiME-4 challenge ${'isolated\_1ch\_track'}$ test set \cite{chime4}. 

\textbf{Implementation Details}
Speech signals are all resampled to 16 kHz. The number of Mel frequencies is set to $F_{\text{mel}}$ = 80 for the frequency range of 0 to 8 kHz. The hidden size is set to $D=192$ (96 for each-directional of full-band BLSTM; 192 for each-directional of sub-band (B)LSTM, but the output dimension is reduced to 192 with a Linear layer for sub-band BLSTM). The full-band plus sub-band LSTMs are repeated 3 times. The number of neighbor frames $N_\text{time}-1$ is set as: 15 past frames for online processing, and plus 15 future frames for offline processing. The number of neighbor frequencies $N_\text{freq}-1$ is set as 5 higher plus 5 lower frequencies. The length of training speech is 3 seconds. We use the Adam optimizer with an initial learning rate of $10^{-4}$, warmup to $10^{-3}$ after 30 epochs, then decreased to $10^{-4}$ with a cosine schedule. The models are all trained for 200 epochs, and the model of last epoch is used as for speech enhancement, while the models of last 10 epochs are averaged as for ASR. 

Some configurations are set differently for speech enhancement and ASR. (i) Speech enhancement. The frame length and hop size of STFT are set to 32 ms and 16 ms, respectively. For joint training of Mel-FullSubNet and Vocos, the loss function and the optimizer setting are remained the same as the original Vocos \cite{Vocos}. (ii) ASR. As for the ASR backends, we employed the REVERB (Transformer) and CHiME-4 (E-Branchformer) recipes in the ESPnet toolkit \cite{espnet}.
To be consistent with ASR, Mel-FullSubNet adopts 32 ms frame length and 8 ms hope size, and the natural logarithm (base of $e$). 

\textbf{Comparison Methods}. We compare with four baseline methods. (i) FullSubNet \cite{Fullsubnet} serves here as a proper baseline for comparing the efficiency of linear-frequency enhancement and Mel-frequency enhancement. 
(ii) Demucs \cite{Defossez2020Real} is an online speech enhancement model that operates directly on raw waveforms. (iii) VoiceFixer \cite{VoiceFixer} also performs Mel-spectrogram enhancement (using a ResUNet) and generates the waveform using a neural vocoder, but is developed only for offline speech enhancement. (iv) DMSEbase (VCTK) \cite{Diffusion} is a followup of VoiceFixer, and uses a more powerfull Mel-spectrogram enhancement network, i.e. the diffusion model \cite{Score}. 
We use the official model checkpoint of baseline methods. Speech is resampled if necessary.  VoiceFixer and DMSEbase are only tested on the CHiME-4 dataset for fair comparison, as the proposed model and these models are all trained not using DNS and Reverb data. 
According to the sampling rate and STFT settings of VoiceFixer and DMSEbase, we retrained the ASR models to allow the enhanced Mel-spectrogram of them directly feeding to the ASR models. 

\textbf{Evaluation Metrics.}
Speech enhancement performance is evaluated with Perceptual Evaluation of Speech Quality (PESQ) \cite{PESQ} and DNSMOS P.835 \cite{Dnsmos}, and we only report the overall audio quality of DNSMOS.
Word Error Rate (WER) is used to evaluate the ASR performance.

\vspace{-0.2cm}
\subsection{Speech Enhancement Results}

Table~\ref{tab:Denoising1} presents the speech enhancement results evaluated on Reverb and DNS, and  Table~\ref{tab:Denoising2} on CHiME-4 and also the model size and computational complexity. 
The real-time factor (RTF) is also measured on a platform equipped with Intel(R) Xeon(R) Platinum 8358 CPU of 2.60 GHz. PESQ is only measured on datasets with available clean reference signals, including the simulated data (simu.) of REVERB and CHiME-4, and the without reverb data (w/o) of DNS. 

\textbf{Performance} Compared to linear-frequency speech enhancement methods, i.e. Demucs and FullSubNet, PESQ scores of the proposed Mel-FullSubNet+Vocos are higher on Reverb and CHiME-4, but lower on DNS. By contrast, the proposed Mel-FullSubNet+Vocos achieves much larger DNSMOS scores on all the three datasets. This phenomenon is consistent with the one found in the Voicefixer paper \cite{VoiceFixer} that Mel-spectrogram enhancement is helpful for improving speech perceptual quality but not for PESQ. 
Transforming the enhanced Mel-spectrogram with GAN-based neural vocoder can achieve high quality enhanced speech, but not necessarily lead to  enhanced speech being close to the reference speech.

\begin{table}[t]
\renewcommand\arraystretch{1.6}
\tabcolsep0.045in
\caption{\footnotesize{Speech enhancement results on REVERB and DNS.}}
\label{tab:Denoising1}
 \centering 
\scalebox{0.87}{
\begin{tabular}{c|c|ccc|ccc}
\hline
\multirow{3}{*}{\textbf{Setting}} & \multirow{3}{*}{\textbf{Model}} & \multicolumn{3}{c|}{\textbf{REVERB}}      & \multicolumn{3}{c}{\textbf{DNS}}       \\ \cline{3-8} 
                                  &                                 & PESQ        & \multicolumn{2}{c|}{DNSMOS} & PESQ      & \multicolumn{2}{c}{DNSMOS} \\ \cline{3-8} 
                                  &                                 & simu.       & simu.         & real        & w/o       & w/           & w/o         \\ \hline
                                  & unpro.                          & 1.22        & 1.82          & 1.66        & 1.58      & 1.39         & 2.48        \\ \hline
\multirow{3}{*}{online}           & Demucs \cite{Defossez2020Real}                         & -           & -             & -           & 2.63      & 2.55         & 3.30        \\
                                  & FullSubNet \cite{Fullsubnet}                       & 2.09        & 2.92          & 2.80        & 2.59      & 2.47         & 3.05        \\
                                  & \textbf{Mel-FullSubNet}            & 2.16        & 3.25          & 3.28        & 2.38      & 3.23         & 3.37        \\ \hline
\multirow{2}{*}{offline}          & FullSubNet \cite{Fullsubnet}                      & 2.19        & 2.90          & 2.75        & 2.76      & 2.47         & 3.06        \\
                                  & \textbf{Mel-FullSubNet}            & 2.29        & 3.27          & 3.31        & 2.63      & 3.28         & 3.37        \\ \hline
\end{tabular}}
\vspace{-0.2cm}
\end{table}

\begin{table}[t]
\renewcommand\arraystretch{1.6}
\tabcolsep0.025in
\footnotesize
\caption{\footnotesize{Speech enhancement results on CHIME-4, and the model size and computational complexity of different methods. For Mel-FullSubNet, VoiceFixer and DMSEbase, and \#Param and MACs  of speech enhancement network / neural vocoder are given in parenthesis.}}
\label{tab:Denoising2}
 \centering 
\scalebox{0.87}{
\begin{tabular}{cccccccc}
\hline
\multicolumn{1}{c|}{\multirow{3}{*}{\textbf{Setting}}} & \multicolumn{1}{c|}{\multirow{3}{*}{\textbf{Model}}} & \multicolumn{3}{c|}{\textbf{CHIIME-4}}    & \multirow{3}{*}{\textbf{\begin{tabular}[c]{@{}c@{}}\#Param\\ (M)\end{tabular}}} & \multirow{3}{*}{\textbf{\begin{tabular}[c]{@{}c@{}}MACs\\ (G/s)\end{tabular}}} & \multirow{3}{*}{\textbf{RTF}} \\ \cline{3-5}
\multicolumn{1}{c|}{}                                  & \multicolumn{1}{c|}{}                                & PESQ  & \multicolumn{2}{c|}{DNSMOS}       &                                      &                                                                                &                               \\ 
\multicolumn{1}{c|}{}                                  & \multicolumn{1}{c|}{}                                & simu. & simu. & \multicolumn{1}{c|}{real} &                                      &                                                                                &                               \\ \hline
\multicolumn{1}{c|}{}                                  & \multicolumn{1}{c|}{unpro.}                          & 1.20  & 1.85  & \multicolumn{1}{c|}{1.30} & -                                    & -                                                                              & -                             \\ \hline
\multicolumn{1}{c|}{\multirow{3}{*}{online}}           & \multicolumn{1}{c|}{Demucs \cite{Defossez2020Real}}                          & 1.61  & 2.98  & \multicolumn{1}{c|}{2.89} & 39.1                                 & 30.62                                                                          & 0.98                          \\
\multicolumn{1}{c|}{}                                  & \multicolumn{1}{c|}{FullSubNet \cite{Fullsubnet} }                      & 1.80  & 2.61  & \multicolumn{1}{c|}{2.39} & 5.6                                  & 30.72                                                                          & 0.66                        \\
\multicolumn{1}{c|}{}                                  & \multicolumn{1}{c|}{\textbf{Mel-FullSubNet}}            & 1.87  & 3.22  & \multicolumn{1}{c|}{3.10} & 15.4(2.2/13.2)                       & 11.2(11.2/0.03)                                                                 & 0.37                          \\ \hline
\multicolumn{1}{c|}{\multirow{4}{*}{offline}}            & \multicolumn{1}{c|}{FullSubNet \cite{Fullsubnet}}                      & 1.87  & 2.62  &  \multicolumn{1}{c|}{2.38} & 14.6                                     & 81.29                                                                          & 1.67                          \\
\multicolumn{1}{c|}{} & \multicolumn{1}{c|}{VoiceFixer \cite{VoiceFixer}}                      & 1.54  & 3.09  & \multicolumn{1}{c|}{2.99} & 104.3(70.4/33.9)                     & 83.6(13.3/70.3)                                                                & 1.63                         \\
\multicolumn{1}{c|}{}                                  & \multicolumn{1}{c|}{DMSEbase  \cite{Diffusion}}                 & 1.48  & 3.28  & \multicolumn{1}{c|}{3.13} & 72.7(58.8/13.9)                      & 414.1(383.6/30.5)                                                              & 9.92                         \\
\multicolumn{1}{c|}{}                                  & \multicolumn{1}{c|}{\textbf{Mel-FullSubNet}}            & 2.02  & 3.29  & \multicolumn{1}{c|}{3.20} & 16.5(3.3/13.2)                                      & 15.7(15.7/0.03)                                                                          & 0.55                         \\ \hline 
\end{tabular}}
\vspace{-0.3cm}
\end{table}

Comprared to FullSubNet, Mel-FullSubNet can generalize better to the unseen CHiME-4 dataset, as the DNSMOS scores does not noticeably degrade from Reverb and DNS to CHiME-4. This also attributes to the better generalization capability of low-dimensional-space learning of Mel-spectrogram enhancement.  
Compared to other Mel-sepectrogram enhancement methods, i.e. VoiceFixer and DMSEbase, the proposed Mel-FullSubNet+Vocos achieves much larger PESQ scores and slightly better DNSMOS scores. The larger PESQ scores reflect that the enhanced speech of Mel-FullSubNet+Vocos are more perceptually closer to the reference speech. 

\textbf{Model size and computational complexity} Compared to FullSubNet, Mel-FullSubNet+Vocos has a larger model size due to the use of Vocos, but much smaller MACs as Mel-FullSubNet processes much less frequencies than FullSubNet (80 versus 256). VoiceFixer and DMSEbase have very large \#Param and MACs. Especially DMSEbase uses the computational expensive diffusion model. 
Overall, the proposed Mel-FullSubNet+Vocos has the lowest RTF. 

\begin{table}[]
\centering
\caption{\footnotesize{Word error rate (WER,\%) on {REVERB}.}}
\label{tab:asr_reverb}
\renewcommand\arraystretch{1.2}
\footnotesize
\setlength{\tabcolsep}{3mm}{
\begin{tabular}{c|cc|cc}
\hline
\multicolumn{1}{c|}{\multirow{2}{*}{\textbf{SE Method}}} & \multicolumn{2}{c|}{simu} & \multicolumn{2}{c}{real} \\ \cline{2-5} 
\multicolumn{1}{c|}{} & near  & far & near & far \\ \hline
unproc. & 3.7 & 4.7  & 6.0 & 6.3\\
\textcolor{gray}{8ch WPE + Beamformit}  & \textcolor{gray}{3.5} & \textcolor{gray}{3.7} & \textcolor{gray}{3.6} & \textcolor{gray}{4.4}\\
1ch WPE \cite{Nakatani2010WPE} & 3.7 & 4.6 & 5.5 & 5.8 \\
FullSubNet \cite{Fullsubnet} & 3.9 & 4.7 & 5.7 & 6.2\\
\textbf{Mel-FullSubNet (prop.)} & \textbf{3.6} & \textbf{4.0} & \textbf{4.2} & \textbf{4.0} \\ \hline
\end{tabular}}
\vspace{-0.2cm}
\end{table}

\begin{table}[]
\centering
\caption{\footnotesize{Word error rate (WER,\%) on {CHiME-4}. `Time' and `Mel' means performing ASR using the time-domain enhanced speech and the enhanced Mel-spetrogram, respectively.} }
\label{tab:asr_chime4}
\renewcommand\arraystretch{1.2}
\setlength{\tabcolsep}{1mm}
\footnotesize
\begin{tabular}{c|ccc|ccc}
\hline
\multicolumn{1}{c|}{\multirow{3}{*}{\textbf{SE Method}}} & \multicolumn{3}{c|}{simu} & \multicolumn{3}{c}{real} \\ \cline{2-7}
\multicolumn{1}{c|}{} & \multirow{2}{*}{unproc.} & \multicolumn{2}{c|}{enhanced} & \multirow{2}{*}{unproc.} & \multicolumn{2}{c}{enhanced} \\
\multicolumn{1}{c|}{} &   & Time  & \multicolumn{1}{l|}{Mel} & & Time  & \multicolumn{1}{l}{Mel} \\ \hline
\textcolor{gray}{5ch Beamformit} & \textcolor{gray}{15.3}  & \textcolor{gray}{13.0} & -  & \textcolor{gray}{13.1} & \textcolor{gray}{10.8} & -  \\
FullSubNet \cite{Fullsubnet} & 15.3  & 24.2 & -  & 13.1  & 18.9 & - \\
VoiceFixer \cite{VoiceFixer} & 15.6 & 27.6 & 24.3 & 13.5 & 30.6 & 26.4  \\
DMSEbase \cite{Diffusion} & 17.1 & 36.7 & 35.2  & 16.0 & 42.6 & 38.6 \\
\textbf{Mel-FullSubNet (prop.)} & 15.3 & 21.6 & \textbf{13.0} & 13.1 & 17.4  & \textbf{10.4}  \\ \hline
\end{tabular}
\vspace{-0.3cm}
\end{table}

\subsection{ASR Results}

Table~\ref{tab:asr_reverb} presents the ASR performance on REVERB. The 1ch WPE \cite{Nakatani2010WPE} and 8ch WPE+Bemformit implemented in ESPnet are also compared. FullSubNet achieves similar WER as unprocessed speech, which is consistent with the common finding in the field \cite{iwamoto2022} that single-channel speech enhancement networks usually do not help for ASR due to the speech artifacts caused by the networks. In contrast, the proposed Mel-FullSubNet achieves much better performance, which is even close to the performance of 8ch WPE+Beamformit. This testifies our assertion that directly enhancing Mel-spectrogram is better than first enhancing linear-frequency spectrogram and then computing Mel-spectrogram for ASR. 


Table~\ref{tab:asr_chime4} shows the ASR performance on CHiME-4. The 5ch Beamformit method implemented in ESPnet is also compared. Again, despite the proposed Mel-FullSubNet is trained using other datasets, it still largely improves the ASR performance, and its performance is even comparable to the one of 5ch Beamformit. 
As described above, we retrained the ASR models for VoiceFixer and DMSEbase, thus their performance on unprocessed speech are different. The enhanced Mel-spectrogram of VoiceFixer and DMSEbase do not improve the ASR performance, which means their enhanced Mel-spectrograms have large artifacts. 
For Mel-FullSubNet, VoiceFixer and DMSEbase, the enhanced Mel-spectrograms all achieve better ASR performance than their waveforms generated by neural vocoders, which further verifies that Mel-spectrogram enhancement is more suitable for robust ASR. 

\section{Conclusion}

This work proposes the Mel-FullSubNet for single-channel speech enhancement. By combining the strong full-band/sub-band fusion network and the strategy of Mel-spectrogram enhancement plus neural vocodor, Mel-FullSubSet achieves a large improvement of speech quality. In addition, the enhanced Mel-spectrogram can be directly used for robust ASR. By comparing with other Mel-spectrogram enhancement networks, the proposed full-band/sub-band fusion network may lead to less speech artifacts, and thus is more ASR friendly. Overall, Mel-spectrogram enhancement with full-band/sub-band fusion network as a whole may open a new door for ASR-oriented single-channel speech enhancement.


\section*{References}

\def\refname{}

\end{document}